\documentclass[aps, prd, twocolumn, superscriptaddress, nofootinbib]{revtex4}
\usepackage{graphicx}% Include figure files
\usepackage{dcolumn}% Align table columns on decimal point
\usepackage{amssymb}% outlined letters
\usepackage{amsmath,amssymb,amsfonts}

\begin{document}

%Macros
\newcommand{\Eq}[1]{\mbox{Eq. (\ref{eqn:#1})}}  
\newcommand{\Fig}[1]{\mbox{Fig. \ref{fig:#1}}}
\newcommand{\Sec}[1]{\mbox{Sec. \ref{sec:#1}}}

\newcommand{\PHI}{\phi}
\newcommand{\PhiN}{\Phi^{\mathrm{N}}}
\newcommand{\vect}[1]{\mathbf{#1}}
\newcommand{\Del}{\nabla}
\newcommand{\unit}[1]{\;\mathrm{#1}}
\newcommand{\x}{\vect{x}}
\newcommand{\y}{\vect{y}}
\newcommand{\p}{\vect{p}}
\newcommand{\ScS}{\scriptstyle}
\newcommand{\ScScS}{\scriptscriptstyle}
\newcommand{\xplus}[1]{\vect{x}\!\ScScS{+}\!\ScS\vect{#1}}
\newcommand{\xminus}[1]{\vect{x}\!\ScScS{-}\!\ScS\vect{#1}}
\newcommand{\diff}{\mathrm{d}}
\newcommand{\mk}{{\mathbf k}}
\newcommand{\ep}{\epsilon}

\newcommand{\be}{\begin{equation}}
\newcommand{\ee}{\end{equation}}
\newcommand{\bea}{\begin{eqnarray}}
\newcommand{\eea}{\end{eqnarray}}
\newcommand{\vu}{{\mathbf u}}
\newcommand{\ve}{{\mathbf e}}
\newcommand{\vn}{{\mathbf n}}
\newcommand{\vk}{{\mathbf k}}
\newcommand{\vz}{{\mathbf z}}
\newcommand{\vx}{{\mathbf x}}
\def\dup{\;\raise1.0pt\hbox{$'$}\hskip-6pt\partial\;}
\def\ddn{\;\overline{\raise1.0pt\hbox{$'$}\hskip-6pt\partial}\;}

%=====================================================================
%=====================================================================
%=====================================================================

\title{Squeezing of scalar and tensor primordial perturbations generated by modified dispersion relations}

\newcommand{\addressImperial}{Theoretical Physics Group, The Blackett Laboratory, Imperial College, Prince Consort Rd., London, SW7 2BZ, United Kingdom}
\newcommand{\addressRoma}{Dipartimento di Fisica, Universit\`a di Roma ``La Sapienza'', P.le A. Moro 2, 00185 Roma, Italy}
\newcommand{\addressRadboud}{Radboud University, Institute for Mathematics, Astrophysics and Particle Physics, Heyendaalseweg 135, NL-6525 AJ Nijmegen, The Netherlands}

\author{Giulia Gubitosi}
\affiliation{\addressRadboud}\affiliation{\addressRoma}
\author{Jo\~{a}o Magueijo}
\affiliation{\addressImperial}

\date{\today}

\begin{abstract}
In recent work we analyzed the evolution of primordial perturbations satisfying Planck-scale-modified dispersion relations and showed that there is no cosmological ``squeezing'' in the critical model that produces perturbations with a scale invariant spectrum. Nevertheless, the perturbations reenter the horizon as standing waves with the correct temporal phase because of the late-time decay of the momentum mode. 
Here we shed light on the absence of primordial squeezing by re-examining the problem in the dual rainbow frame, where $c$ is set to 1, shifting the varying $c$ effects elsewhere. In this frame gravity switches off at sub-Planckian wavelengths, so that the fluctuations behave as if they were in Minkowski spacetime. This is ultimately why they are not squeezed. However, away from the critical model squeezing does occur if the fluctuations spectrum is red, as is the case for scalar perturbations. Should the primordial gravity waves have a blue spectrum, we predict that they might not reenter the horizon as standing waves, because the momentum mode would be enhanced in the primordial phase. 
\end{abstract}


\maketitle

\section{Introduction}

Squeezing of quantum fluctuations has long been paraded as a key feature of inflation (e.g.~\cite{grish, Grishchuk:1993zd, starob96}). In a recent paper~\cite{us}, however, we showed that once one focuses on the end-product of squeezing, just about any model complies with the observational constraints. All that is observationally needed is that at late-time horizon reentry the fluctuations form standing waves with the correct temporal phase \cite{Dodelson:2003ip}. As shown in~\cite{us} this does not require that in the primordial phase there was ``squeezing'' (seen as the suppression of the momentum mode with respect to the momentum-free mode).
It is sufficient for the  momentum mode not to be overwhelmingly large at the end of the primordial phase, so that its decay during the standard Big Bang epoch leaves a dominant momentum-free mode at late times, producing the required standing wave.  Thus, inflation and other models where fluctuations are squeezed (such as bimetric VSL models \cite{Jprl, Jprd}) are actually ``overkill'', or surplus to requirement. 

This remark is particularly pertinent for some models based on modified dispersion relations (MDR) \cite{DSRJM, Mukohyama:2009gg,DSR-dimred, DSR-rainbow, Amelino-Camelia:2013gna}, specifically  the ``critical'' MDR model. This model is  characterized by the dispersion relation found in Horava-Lifshitz theory, and is known to produce an exactly scale-invariant spectrum of primordial fluctuations. The critical model does not squeeze the primordial fluctuations, injecting equal amounts of momentum and momentum-free modes into the standard radiation epoch \cite{us}. As shown in~\cite{us} this is phenomenologically acceptable. Yet,  one may wonder about the physical origin of this result, particularly as it is an oddity with respect to almost every other scenario. 

In this paper we shed light on this matter, re-examining the model in the dual frame, where the speed of light $c$ is set to one as a result of a wavelength-dependent redefinition of time. It was previously suggested that in this dual frame, driven by rainbow gravity, ``gravity switches off'', or matter becomes ``conformally coupled to gravity'' \cite{essay, DSR-rainbow}. In Sections~\ref{dual}, 
\ref{squeezing}, 
\ref{critical} 
we show that the former description is more precise. In a radiation dominated Universe (i.e. with conformal coupling to gravity) there is squeezing; in Minkowski spacetime there is not. The fluctuations in the critical MDR model behave for the purposes of squeezing just as if they were in Minkowski spacetime; therefore for all practical purposes gravity is ``switched off''. In this way this paper clarifies both the result on squeezing and the nature of gravity for the critical MDR model.

Beyond the critical model, we know~\cite{us} that for MDR models with a red spectrum of perturbations squeezing does occur. Thus,  the observed slightly red scalar perturbations must behave similarly to inflation, even if produced by MDR. In Section~\ref{gw} we show that substantial novelties with regards to inflation might arise for tensor perturbations. The MDR for gravitons can in principle be different from that for scalar perturbations; in particular it can produce a blue spectrum. In that case not only would there be no squeezing, but the momentum mode could grow large enough (compared to the momentum-free mode)
that the evolution in the standard radiation phase would not be able to suppress it before horizon reentry.

We find that this possibility depends solely on the background equation of state during the MDR phase.
If $w<1$ then the usual standing waves are formed. If $w>1$ tensor perturbations form standing waves with a cosine temporal phase (complementary to the sine phase of scalar perturbations, responsible for the observed position of the Doppler peaks). 
For $w=1$  tensor perturbations would form traveling waves upon horizon reentry.

\section{MDR frame and its dual frame}\label{dual}
The theory of cosmological fluctuations for MDR models was first developed~\cite{DSRJM,DSR-dimred} in a frame where the existence of MDRs is made explicit and Einstein gravity is valid in a well-defined sense. This frame is thus called the MDR or Einstein frame. In this frame the second order action for the fluctuations is:
\bea
S_2&=&\int d^3 k d\eta \, z^2[\zeta'^2 -c ^2 k^2 \zeta^2]
\eea
with $z= a$ and $c$ given by: 
\be
c= \left(\frac{\lambda k}{a(\eta)}\right)^{\gamma}\,,\label{eq:speedgamma}
\ee
where $\lambda^{-1}$ is the UV scale and $\gamma$ is a dimensionless parameter.  The critical model with $\gamma=2$ reproduces the dispersion relation found in the Horava-Lifshitz critical model and leads to perturbations that are scale invariant already inside the horizon \cite{DSR-dimred}. 

As usual, one sets $\zeta=y/a$ to obtain the dynamical equation:
\be
y''+\left(c^2k^2-\frac{a''}{a}\right)y=0.
\ee
The conjugate momentum to $y$ is given by 
\be\label{mom1}
p=y'-\frac{a'}{a}y=a\zeta'.
\ee
In general the positive and negative frequencies can be identified from:
\bea
y(\vk,\eta)&=&\frac{c(\vk,\eta)+c^\dagger (-\vk,\eta)}{\sqrt{2\omega}} \label{eq:yvsc}\\
p(\vk,\eta)&=&-i \sqrt{\frac{\omega}{2}}(c(\vk,\eta)- c^\dagger (-\vk,\eta)).\label{eq:pvsc}
\eea
An analysis of the squeezing parameter $s$ in this frame was carried out in~\cite{us}, with the result that for $\gamma=2$ no squeezing occurs: $s=0$. One may also compute~\cite{us} the parameter $\sigma$ that measures the ratio of the momentum-free mode and the momentum mode (equivalent to $s$ asymptotically for many models, including inflation) with the result:
\be\label{sigma}
\sigma=\frac{\omega^2 |y|^2}{4|p|^2}\sim1.
\ee
Hence no suppression of the momentum mode over the momentum-free mode occurs in the MDR phase; but also neither is there an enhancement of this mode, as would be the case for a collapsing universe (see~\cite{us}).

As explained in~\cite{DSR-rainbow}, a dual frame, with a constant $c$, may be obtained by defining a new time variable: 
\be\label{tau}
\tau=\int c d\eta
\ee
(in what follows we shall use a tilde to denote quantities as measured in the dual frame).
This is a ``rainbow frame'' because the new time is also $k$ dependent (i.e.: $\tau=\tau(\eta,k)$).
In such a frame $\tilde c=1$, however the non-trivial effects of what in the MDR frame is a varying-$c$ are shifted elsewhere. 
In the dual frame Einstein's gravity is no longer valid, and is replaced by rainbow gravity~\cite{DSR-rainbow}. Setting $\tilde\zeta=\zeta$ it is straightforward to show that: 
\bea
S_2&=&\int d^3 k d\tau\, \tilde z^2[\dot{\tilde \zeta}^2 - k^2 \tilde \zeta^2]\\
\tilde z &=& a\sqrt{c}
\eea
where a dot denotes derivative with respect to $\tau$. The equation of motion associated with this action is:
\be\label{eom1}
\ddot {\tilde y} + \left(k^2-\frac{\ddot{\tilde z}}{\tilde z}\right)
\tilde y=0,
\ee
with $\tilde \zeta=\tilde y/\tilde z$, as usual. 
It is remarkable that for $\gamma=2$ we obtain a time independent parameter $\tilde z$ controlling the effects of gravity upon the fluctuations:
\be
\tilde z=\lambda k.
\ee
Although the $k$ dependence in $\tilde z$ signals a non-local relation between $\tilde \zeta$ and $\tilde y$, its lack of time-dependence has an important implication. It marks a decoupling of the fluctuations from gravity (or the ``switching off of gravity'', as speculated in~\cite{essay}). In fact, for $\gamma=2$ the fluctuations live as if they were in Minkowski spacetime, with dynamical equation:
\be\label{eom}
\ddot {\tilde y} + k^2\tilde y=0,
\ee
since the term in $\ddot{\tilde z}/\tilde z$ now vanishes. Moreover, for $\gamma=2$ the conjugate momentum to $\tilde y$ is simply:
\be\label{mom}
\tilde p=\dot{\tilde y}.
\ee
This is to be contrasted with the result for a radiation dominated universe (where a term in $a'/a$ subsists in the relation between $y$ and its conjugate momentum), with implications to be made apparent soon.

\section{A squeezing dictionary between frames}\label{squeezing}
We now provide a dictionary between frames for the quantities appearing in the squeezed state formalism, in order to facilitate a re-examination of the findings in~\cite{us} from the perspective of the dual frame. In providing this dictionary we shall keep $\gamma$ general.

Taking $\tilde \zeta=\zeta$ as a starting point, and setting $\zeta=y/z$ and $\tilde \zeta= \tilde y/\tilde z$ as usual,  we have at once that:
\be
\tilde y=y\frac{\tilde z}{z}=y\sqrt{c}.
\ee
In addition, using  (\ref{mom1}), we find the relation between the momenta:
\be
\tilde p=\tilde z \dot\zeta=p\frac{\tilde z}{z}\frac{1}{c}=\frac{p}{\sqrt{c}}.
\ee
In view of these relations, and considering that $\omega=c k$,
comparing (\ref{eq:yvsc}) and (\ref{eq:pvsc}) (valid in the Einstein/MDR frame) with their homologous in the  dual frame:
\bea
\tilde y(\vk,\tau)&=&\frac{\tilde c(\vk,\tau)+\tilde c^\dagger (-\vk,\tau)}{\sqrt{2 k}} \label{eq:yvsc1}\\
\tilde p(\vk,\tau)&=&-i \sqrt{\frac{ k}{2}}(\tilde c(\vk,\tau)- \tilde c^\dagger (-\vk,\tau))\label{eq:pvsc1}
\eea
we can conclude that:
\bea
c(\vk,\eta)&=&\tilde c(\vk,\tau)\\
c^\dagger (-\vk,\eta)&=&\tilde c^\dagger (-\vk,\tau).
\eea
Therefore the Bogolubov transformations in the MDR frame:
\bea
c(\vk,\eta)&=&u_\vk(\eta)c_0(\vk)+  v_\vk(\eta)c^\dagger _0(-\vk) \label{eq:cEvolution}\\
c^\dagger (-\vk,\eta)&=&v^\star_\vk(\eta)c_0(\vk) + u^\star _\vk(\eta)c^\dagger _0(-\vk)\label{eq:cdagEvolution}
\eea
are exactly mimicked in the dual frame with $\tilde u_\vk=u_\vk$ and $\tilde v_\vk=v_\vk$, with the consequence that when we parameterise 
\bea
u_{k}(\eta)&=&e^{-i\theta_{k}(\eta)} \cosh(r_{k}(\eta))\label{uexp}\\
v_{k}(\eta)&=&e^{i(\theta_{k}(\eta)+2 \phi_{k}(\eta))} \sinh(r_{k}(\eta))\label{vexp},
\eea
and likewise for the dual frame, we find that squeezing parameter and angles are the same in both frames. 
Specifically: 
\be
\tilde s_k=s_k
\ee
(where $s_\vk(\eta)=|v_\vk(\eta)|^2$, and likewise for the dual frame). 

It is also straightforward to see that the parameter $\sigma$ introduced in~\cite{us} and defined in (\ref{sigma}) is invariant under frame transformation: 
\be
\tilde \sigma=\frac{|\tilde y|^2 k^2}{4 |\tilde p|^2}= \frac{|y|^2 \omega^2}{4 |p|^2}=\sigma.
\ee
The relative strengths of the momentum mode and the momentum-free mode are therefore the same, once we include the different value of $c$ in the two frames. Therefore arguments laid out in the dual frame transpose directly to the Einstein/MDR frame. This will help us understand better the result found in~\cite{us}.

\section{Explaining the solutions in the MDR frame for the critical model}\label{critical}
It should be obvious from Eqs.~(\ref{eom}) and (\ref{mom}) that for $\gamma=2$ the fluctuations behave as if they were living in Minkowski spacetime. This is to be distinguished from fluctuations living in a radiation dominated universe with Einstein gravity. In fact in that case  Eq.~(\ref{eom}) is valid as well (since the term in $a''/a$ in the equation of motion vanishes, due to conformal coupling). However, the effects of expansion are still present in the definition of the conjugate momentum, $p=y'-y/\eta$. This leads to squeezing, as explained in~\cite{us}. In the present case things are quite different in this respect, as we now see.

The most general solution to (\ref{eom}) is:
\be\label{ysoldual}
\tilde y= \frac{1}{\sqrt{2k}}(\tilde c_0(\vk)e^{-ik\tau}+ \tilde c_0^\dagger (-\vk)e^{ik\tau})
\ee
with conjugate momentum:
\be
\tilde p =\dot{\tilde y}=- i\sqrt{\frac{k}{2}}(\tilde c_0(\vk)e^{-ik\tau}- \tilde c_0^\dagger (-\vk)e^{ik\tau}).
\ee
Comparing with (\ref{eq:yvsc1}) and (\ref{eq:pvsc1}) we therefore find:
\bea
\tilde c(\vk,\eta)&=&\tilde c_0(\vk)e^{-ik\tau}\\
\tilde c(-\vk,\eta)&=& \tilde c_0^\dagger (-\vk)e^{ik\tau}
\eea
so that $v_\vk=0$ and there is no squeezing at all,
\be
\tilde s=0,
\ee
as expected in Minkowski spacetime.  Also it is obvious that for vacuum fluctuations at late times we have:
\be
\tilde \sigma\sim 1.
\ee
This explains the results found in~\cite{us}. Since $\tilde s=s$ and $\tilde\sigma=\sigma$, both squeezing and momentum suppression can be equally understood in both frames. In the dual frame the fluctuations behave as if they were living in Minkowski spacetime. Thus, there is no squeezing or momentum suppression, just as was found in~\cite{us}. 

It also explains why the solutions found in~\cite{us} are so simple for $\gamma=2$. They are in fact  harmonic oscillations transposed to the MDR frame (note that Hankel functions of order $\nu=1/2$ are just harmonic oscillations). To make this explicit note that if: 
\be
a\propto \eta^m
\ee
with 
\be
m= \frac{2}{1+3w} \label{mw}
\ee
then the relation between the two times given by (\ref{tau}) is:
\be\label{taueta}
\tau =\frac{-\lambda^2 k^2}{(2m-1)\eta ^{2m-1}}.
\ee
The minus sign represents the fact that growing $\eta>0$ maps into growing $\tau<0$ if 
$-1/3<w\le1$ (or $m\ge 1/2$), needed for solving the horizon problem (see~\cite{us}). Inserting \eqref{taueta} into (\ref{ysoldual}) we recognize the solutions found in~\cite{us}, with amplitudes $c(\vk)=\tilde c(\vk)$, with the sign of the exponentials flipped (as explained there, this is needed so that $\vk$ points to the actual direction of propagation).

\section{Non-critical models and gravity waves}\label{gw}
It was shown in~\cite{us} that, should $\gamma\neq 2$, then $\sigma$ changes in time and therefore will not be of order 1 at the end of the primordial MDR phase, as is the case for $\gamma=2$. However, the observed red spectrum of scalar fluctuations suggests $\gamma\ <2$ (see \cite{DSR-dimred}), for which it was shown that $\sigma$ {\it increases} in time,
much as in inflation. 
Therefore no new constraints upon the theory are obtained. The situation would potentially be different if the spectrum could be blue (and $\gamma >2$), because then $\sigma$ could {\it decrease} during the MDR phase to the point where its growth in the radiation epoch might not be enough to ensure the production of the correct standing waves at horizon reentry. This cannot happen for scalar perturbations since their spectrum has been observed to be red. For tensor perturbations, however, it remains a possibility, with implications explored in this Section.

As was stated in \cite{DSR-dimred} much of the discussion for scalar modes in MDR models can be replicated for tensor modes, assuming a dispersion relation of the same form, but with possibly different values for the parameters $\gamma$ and $\lambda$ (cf. Eq. \eqref{eq:speedgamma}). We should just add $S$ and $T$ labels to all variables, to distinguish the two types of fluctuations. The fact that $\lambda_S$ could be different from $\lambda_T$ leads to different amplitudes (and thus controls the observable tensor to scalar ratio $r$  \cite{DSR-dimred}). It could also be that the exponents are different, $\gamma_S\neq \gamma_T$, with the implication that $n_S\neq n_T$. Given that no primordial tensor modes have yet been observed, we can speculate on the implications for the squeezing of tensor modes, should they have a blue spectrum. Could it be that for a given range of blue spectra the gravity waves reenter the horizon as travelling waves, or even as standing waves with the complementary phase to the one observed for scalar modes?

In order to answer this question we note that during the MDR phase we have \cite{us}:
\be\label{sigmaDSR}
\sigma_T \propto
a^{4-2\gamma_{T}},
\ee
so that for $\gamma_T>2$ it decreases in time starting from a value of order 1. We should therefore evaluate 
 the evolution of $\sigma_T$ in the radiation epoch, should it be fed a very small $\sigma_T$ from the primordial phase. Until $\sigma_T\sim 1$ we have $y_T\approx B\eta/a$ and $p_T\approx B'/a$, so that
\be\label{sigmarad1}
\sigma_T\propto \eta^2
\ee
(this is the same result as for a contracting radiation dominated universe, since the evolution has $\sigma\ll 1$ throughout). Once $\sigma\sim 1$ (if this ever happens), then $y\approx A $ and $p\approx B'/a$, and we have:
\be\label{sigmarad2}
\sigma_T\propto a^4\propto \eta^4.
\ee
This is consistent with the formula relating $\sigma_0$ and $\Sigma$ in \cite{us}, valid when  $\sigma_0$ is not much smaller than 1. In our problem the two regimes (Eq.~(\ref{sigmarad1}) and Eq.~(\ref{sigmarad2})) may need to be considered.

We now need to work out the condition for the decay in $\sigma_T$ in the MDR phase to be severe enough that its growth in the radiation epoch is not sufficient to make it large upon horizon reentry. 
The specific dependence of the scale factor on time depends on both the parameter $\gamma_T$ and the equation of state parameter $w$.
The generalization of (\ref{taueta}), relating the conformal MDR frame time and the rainbow time $\tau$ for a generic $\gamma_T$, is:
\be
\eta\propto (-\tau)^{\frac{-1}{\gamma_T\, m-1}}
\ee
so that the evolution of the scale factor is:
\be
a\propto (-\tau)^{\frac{-m}{\gamma_T\, m-1}},
\ee
for $m$ defined in eq. \eqref{mw} and $\gamma_T \,m>1$. This last condition is needed so that we have inflation in the dual frame and solve the horizon problem. In terms of the equation of state parameter it is equivalent to 
\be\label{horsol}
-\frac{1}{3}<w<\frac{2\gamma_T-1}{3},
\ee
something that was know since \cite{DSR-rainbow}. Combining these results we see that during the MDR phase:
\be\label{sigmatau}
\sigma_{T}\propto (-\tau)^{\frac{-m(4-2\gamma_T)}{\gamma_T \,m-1}}\,.
\ee

The advantage of using $\tau$ instead of $\eta$ to discuss our problem is that the ratio between $|\tau|$ at horizon first crossing and the end of the MDR phase and the ratio of $|\eta|$ at horizon reentry and at the end of the MDR phase is the same. Hence we can find directly the conditions for $\sigma$ not to decrease so much in the MDR phase so that it cannot grow again to be large in the radiation epoch. The condition is simply that the exponent in (\ref{sigmatau}) be smaller than the one in  (\ref{sigmarad1}). The limiting condition (that the exponent is 2) would translate into travelling waves on reentry, since the momentum-free mode and the momentum mode would have  similar amplitudes. This happens for $m=1/2$, that is
\be
w=1
\ee
and any value of $\gamma_T>2$ (see (\ref{horsol})). If $1<w<\frac{2\gamma_T-1}{3}$ the exponent in \eqref{sigmatau} is larger than 2, indicating that the momentum mode is dominant at horizon reentry. Thus  near-standing waves with a cosine temporal phase would reenter the horizon. If $w<1$, on the other hand, the exponent in \eqref{sigmatau} is smaller than 2, so that the growth of the momentum mode  during the MDR phase is more than compensated by  its suppression during the following radiation epoch. In this case near-standing waves with the sine temporal phase would reenter the horizon, similarly to scalar perturbations.

We have demonstrated that the temporal phase of gravity waves at late time horizon reentry only depends on the equation of state parameter for any $\gamma_{T}>2$, that is for any value of the spectral index $n_{T} > 1$. Hence, even if gravity waves were blue, we would not get a direct constraint on $n_T$ from the phenomenology of primordial gravity waves, should these ever be observed. Instead we would constrain the equation of state during the MDR phase. The observation of travelling waves, in particular, would require $w=1$. 
From another perspective, if we were to observe that the position of the Doppler peaks of gravity waves is compatible with a cosine temporal phase at horizon reentry, rather than a sine phase, then MDR models would predict a blue spectrum and $w>1$.

\section{Conclusions}
In this paper we reexamined the status of squeezing of primordial fluctuations produced 
in models with MDR. New insights were obtained from the perspective of the 
dual rainbow frame, where perturbations propagate with constant speed, but time is wavelength-dependent. 
We first focused on the model characterized by the dispersion relation that leads to an exactly  scale-invariant power spectrum ($\omega^{2}\sim k^{6} $). We found that in the rainbow frame perturbations propagate following the same dynamics that they would have in Minkowski spacetime. This happens both at the level of the equations of motion and their solutions (harmonic oscillations) and at the level of the definition of the conjugate momentum to perturbations. Thus the absence of squeezing (which would still be expected if the fluctuations were simply conformally coupled to gravity, as is the case of a radiation dominated universe). Absence of primordial squeezing, however, does not lead to pathological implications. Equal amounts of momentum and momentum-free modes would be injected in the radiation dominated phase. The former, then, decays,  so that 
perturbations reenter the horizon as standing waves with a sine temporal phase, similarly to what happens in inflation. 

In more general models we have $\omega^{2}\sim k^{2(\gamma+1)} $, and $\gamma$ could be different for scalars and tensors.  If $\gamma<2$ then the spectrum is red; thus this would be a more realistic model for scalar perturbations. In that case squeezing does occur, so no further constraints arise. The case $\gamma>2$ is interesting in that during the primoridal phase the momentum mode is enhanced over the momentum-free mode (just like in a contracting universe), possibly leading to new results. However, in this case the spectrum is blue. Therefore this could only be relevant for tensor perturbations, since for them we do not yet have constraints on the spectral index of their power spectrum. 

If the spectrum of tensor fluctuations were to be blue, we find that whether or not perturbations reenter the horizon as standing waves depends on the equation of state during the MDR phase. For $w>1$ one would have standing waves with a cosine phase, while a sine phase would be expected for $w<1$. If $w=1$ one would expect  perturbations to reenter the horizon as travelling waves. This relation between the kind of waves formed by the perturbations and the equation of state is likely to be a unique feature of MDR models, with interesting potential phenomenological implications in the distant future.

\section*{Acknowledgments}
We thank Robert Brandenberger, Carlo Contaldi and
Marco Peloso for discussions related to this paper.  We acknowledge support from the John Templeton Foundation. JM was also supported by an STFC consolidated  grant.

%=====================================================================
%=====================================================================
%=====================================================================

\end{document}